\documentclass[journal]{IEEEtran}
\usepackage{ifpdf}
\usepackage{cite}
\usepackage{algorithmic}
\usepackage[ruled,vlined]{algorithm2e}
\usepackage{subcaption}
\usepackage{graphicx}
\usepackage{titlesec}
\titlespacing*{\section}{0pt}{0.5\baselineskip}{0.5\baselineskip}
\titlespacing*{\subsection}{0pt}{0.5\baselineskip}{0.5\baselineskip}
\usepackage[cmex10]{amsmath}
\usepackage{epstopdf}
\usepackage{array}
\usepackage{flushend}
\usepackage{balance}
\usepackage{eqparbox}

\usepackage{fixltx2e}
\usepackage{color}
\usepackage{float}

\usepackage{stfloats}
\usepackage{url}
\usepackage{cite}
\usepackage{amsthm}

\usepackage{amssymb}
\begin{document}

\title{Alternating Channel Estimation and Prediction for Cell-Free mMIMO with Channel Aging: A Deep Learning Based Scheme }

\setlength{\columnsep}{0.21 in}

\author{Mohanad~Obeed,\thanks{M. Obeed and A. Chaaban are with the School of Engineering,  University of British Columbia, Kelowna V1Y 1V7, BC, Canada (email: mohanad.obeed@ubc.ca, anas.chabaan@ubc.ca) and Y. Al-Eryani is a Baseband System Developer at Ericsson Canada, Ottawa, Canada (email: yasser.al-eryani@ericsson.com).} Yasser Al-Eryani, and Anas Chaaban,~\IEEEmembership{Senior Member,~IEEE,}}

\maketitle

\begin{abstract}
  In large scale dynamic wireless networks, the amount of overhead caused by channel estimation (CE) is becoming one of the main performance bottlenecks. This is due to the large number users whose channels should be estimated, the user mobility, and the rapid channel change caused by the usage of the high-frequency spectrum (e.g. millimeter wave). In this work, we propose a new hybrid channel estimation/prediction (CEP) scheme to reduce overhead in time-division duplex (TDD) wireless cell-free massive multiple-input-multiple-output (mMIMO) systems. The scheme proposes sending a pilot signal from each user only once in a given number (window) of coherence intervals (CIs). Then minimum mean-square error (MMSE) estimation is used to estimate the channel of this CI, while a deep neural network (DNN) is used to predict the channels of the remaining CIs in the window. The DNN exploits the temporal correlation between the consecutive CIs and the received pilot signals to improve the channel prediction accuracy. By doing so, CE overhead is reduced by at least 50 percent at the expense of negligible CE error for practical user mobility settings. Consequently, the proposed CEP scheme improves the spectral efficiency compared to the conventional MMSE CE approach, especially when the number of users is large, which is demonstrated numerically.
\end{abstract}
\IEEEpeerreviewmaketitle
\section{Introduction}
Recently, the concept of a cell-free network architecture has been considered as an efficient practical solution to cope with the massive growth of the number of wireless devices and access points (APs).  
In a cell-free  wireless network, a large number of distributed APs cooperate to serve a large number of  users over a geographical area. As such, cell-free (or cell-less) network architectures are envisioned to increase the coverage and transmission rates in future generation wireless networks~\cite{Cell_Ngo}.   

Acquiring accurate channel state information (CSI) in such networks is important to improve the uplink and downlink transmission performance.
However, with the exponential increase of the number of devices, channel estimation (CE) overhead becomes a limiting factor that significantly decreases the spectral efficiency of the system, especially if these users are mobile or the coherence interval is short. This means that classical pilot-based CE approaches will become inapplicable for cell-free massive MIMO system.       

In the last few years, deep learning (DL) based approaches for wireless communication systems have received escalating research attention due to their ability to provide solutions for highly complex problems of resource allocations, estimation, detection, and coding, among others. For CSI acquisition, deep-learning based approaches have proved considerable improvement compared to traditional approaches. The main idea is that deep neural networks (DNN) fed by a large CSI dataset can be trained to perform CE \cite{dong2019deep}, pilot contamination reduction \cite{kim2018deep}, or channel prediction \cite{jiang2020deep}, by learning spatial and/or temporal correlations from the dataset. 

%% Related work
Several papers considered CE and prediction to reduce the overhead via DL approaches\cite{8482358, Mash_Pruning, zhang2020deep, yuan2020machine}. In \cite{8482358}, the authors developed a DL-based CSI feedback protocol for frequency division duplex (FDD) MIMO systems by combining CsiNet (a DL network to reconstruct the CSI) with a long short-term memory (LSTM) network. They showed that the proposed architecture can significantly improve the recovery quality of compressed CSI with a negligible increase in the feedback overhead. The authors of \cite{Mash_Pruning} employed a DNN to jointly design pilots and estimate the downlink channels, where the DNN aids in reducing the pilot transmission overhead and learning long-term correlations in the channel matrix to improve the CE performance. When the antennas of a massive MIMO system use only 1-bit analog-to-digital converters, the authors of \cite{zhang2020deep} leveraged a DNN and the prior CE observations to learn how to convert the quantized received measurements into channel estimates. 

The aforementioned works focused on designing pilot signals and recovering CSI using DL-based approaches. In this work, we focus on predicting  CSI using temporal correlation, where CSI is estimated using classical methods for part of the time, and predicted using DL models for the rest of the time. Such an approach was studied in 
 \cite{yuan2020machine} which proposed a channel prediction scheme for a time division duplex (TDD) massive MIMO system, employing a convolutional neural network to learn the channel aging pattern and an autoregressive predictor to  predict the channel of several consecutive CIs. The proposed scheme in \cite{yuan2020machine} thus requires estimating the channels of several consecutive CIs to enable learning the channel aging pattern, which is then used to predict the channels for several CIs. However, predicting channels relaying on old CSI would lead to a poor prediction accuracy when channels vary relatively fast. In addition, due to the error propagation of channel predictions, utilizing the previously predicted channels to help in predicting the upcoming ones leads to increasing the mean-square error (MSE) significantly. To solve this problem, we propose an alternating hybrid channel estimation/prediction (CEP) scheme to trade-off between reducing the CE overhead and minimizing the MSE.  

%% Contribution

Specifically, this paper first proposes an alternating estimation/prediction scheme to reduce the CE overhead by at least 50 percent in a TDD cell-free massive MIMO system. In the proposed scheme, a window of two or more CIs is divided into an estimate-and-transmit (ET) CI (first CI) and one or more predict-and-transmit (PT) CIs (remaining CIs). During the ET CI, each user transmits its pilot signal to all APs and each AP uses the received signal to estimate the channels during the ET CI, which are then used to optimize the transmission. Then, during the PT CI, the APs use a pre-trained deep neural network (DNN) to predict the channels of the remaining CIs, which are then used to optimize the transmission during these CIs.  The DNN uses a pre-defined number of  received pilot signals during the previous estimation phases to learn how the channel evolves over time in order to accurately predict the channels during the PT CI. As a result, the input of the DNN is updated periodically with estimated channels, which solves the aforementioned problems of poor prediction and error propagation. 

 The proposed scheme is a distributed scheme since it can be implemented at each AP without the need for exchanging any information between APs. The central processing unit (CPU) can be first used to train the DNN model offline. Then, it distributes the DNN parameters to all APs to implement the CEP independently.  The proposed scheme provides a significant improvement in system throughput over the traditional MMSE approach 
 %, reaching up to $15\%$ improvement 
 for a practical number of APs and users and practical user mobility.    

The rest of the paper is organized as follows. Sec. II presents the system and channel models used in this work. In Sec. III, the baseline approach (MMSE) is presented and discussed. Sec. IV presents the proposed DL-based approach. Detailed performance analysis of the proposed scheme is presented in Sec. V, while Sec. VI concludes this work.
 
\section{System Model}
We consider a TDD system consisting of a massive number of $M$ single-antenna APs distributed over a certain area and serving $K$ single-antenna users simultaneously. The APs need to send information to the users using a scheme that requires CSI.  We consider a fixed CI duration for all users. We consider channels that vary in a correlated manner over time, but their variation is negligible within a CI. We model this scenario with channels that stay fixed during every CI and change between CIs according to a given auto-correlation function (ACF) \cite{yuan2020machine}. Thus, CE must be done periodically to update CSI, which means that the system will alternate between CE and downlink data transmission. We describe this in more detail next, starting with the channel aging model, followed by the CE, and then the downlink transmission. 
%We consider only the downlink transmission to evaluate the performance of the proposed channel estimations. Hence, each CI, which is in a TDD structure, is divided into two blocks: CE and downlink payload phase.  

\subsection{Channel Aging Model}
The channel between user $k$ and AP $m$ during CI $\ell$ is given by
\begin{equation}
    g_{m,k}[\ell]=\sqrt{\beta}h_{m,k}[\ell],
\end{equation}
where $\beta \in \mathbb R$ is the large scale fading and $h_{m,k}[\ell] \in \mathbb C$ is the small scale fading during CI $\ell$.
The main reason of channel aging is the users movement, and this can be characterized by the ACF \cite{ARC, CSI_Cor_Learning_2}. The normalized discrete-time ACF is given by
\begin{equation}
    R[\ell]\triangleq J_0(2\pi f_n |\ell|),
\end{equation}
where $J_0()$ is the zeroth-order Bessel function of the first kind, $\ell$ is the index of the CI, and $f_n=vT_sf_d$ is the normalized Doppler shift with sampling duration $T_s$, maximum Doppler frequency $f_d$, and a number of samples $v$ in a CI.

Using this ACF, the small scale fading channel can be expressed as 
\begin{equation}
\label{hmk}
    h_{m,k}[\ell]=-\sum_{q=1}^Q a_{m,k,q}h_{m,k}[\ell-q]+w[\ell],
\end{equation}
where $w[\ell] \backsim \mathcal{CN}(0,\sigma_w^2)$ is a circularly symmetric complex white Gaussian noise process with zero mean and variance $\sigma_w^2=R[0]+\sum_{q=1}^Q a_{m,k,q}R[-q]$, and $a_{m,k,q}$, $q=1,\ldots, Q,$ are the autoregressive coefficients that are the entries of the vector
\begin{equation}
 \mathbf{a}_{k,m}=-\mathbf{R}^{-1}\mathbf{u}   
\end{equation}
where 
$\mathbf{R}=\begin{bmatrix} R[0]&R[-1]&\ldots&R[1-Q]\\R[1]&R[0]&\ldots&R[2-Q]\\.&.&.&.\\.&.&.&.\\R[Q-1]&R[Q-2]&\ldots&R[0]  \end{bmatrix},\ \ \ \ \ $
and $ \mathbf{u}=[R[1],R[2],\ldots,R[Q]]^T$. Note that $R[0]=1 $.

Equation \eqref{hmk} indicates that the small scale fading channel is a weighted sum of complex white Gaussian random variables, which means that the channel $h_{m,k}[\ell] \backsim \mathcal{CN}(0,\sigma^2_{h_{m,k}})$, where $\sigma^2_{h_{m,k}} = \sum_{j=1}^{\infty} F_j^2\sigma^2_w$ can be found using Green's function \cite{box2015time}, where $F_0=1,$ $F_j=\sum_{q=1}^j a_{k,q}F_{j-q}$ if $j\leq Q, $ and $F_j=\sum_{q=1}^Q a_{k,q}F_{j-q} $ if $j> Q $.
\subsection{CE Phase}
In the classical TDD scheme, each CI is divided into three parts, one is devoted for CE, and the remaining two are used for uplink and downlink data transmission. We describe the CE here.
%Since we evaluate here only the downlink transmission, we assume that each coherence interval consists of CE phase and downlink transmission phase. The extension to consider also the uplink transmission is straightforward, however, due to the space limitation, we consider only the downlink transmission.  

In the CE phase, the users send their assigned pilot signals to the APs. We assume that the assigned pilot signals are all orthogonal and have unit energy, which means that the number of pilot signal is greater than or equal to $K$. Let $\mathbf{q}_k$ be the pilot signal (in vector form) sent by user $k$. The received signal at  AP $m$ in CI $\ell$ is given by
\begin{equation}
\mathbf{y}_m[\ell]=\sqrt{\rho\tau_c}\sum_{k=1}^K g_{m,k}[\ell]\mathbf{q}_k+\mathbf{n}
\end{equation}
where $\rho$ is the transmit power during the CE phase, $\tau_c$ is the number of symbols of the pilot signal,  and $\mathbf{n}$ is circularly symmetric  additive white Gaussian noise with zero mean and identity covariance matrix. 
%Note that $g_{m,k}[\ell]=\sqrt{\beta_{m,k}}h_{m,k}[\ell]$, where $\beta_{m,k}$ and $h_{m,k}[\ell]$ are the large-scale and small-scale fading. 
By projecting $\mathbf{y}_m[\ell]$ onto $\mathbf{q}_k$, AP $m$ obtains
\begin{equation}
    \bar{y}_{m,k}[\ell]=\mathbf{q}_k^H\mathbf{y}_m[\ell]=\sqrt{\rho\tau_c} g_{m,k}[\ell]+\mathbf{q}_k^H\mathbf n
\end{equation}
The MMSE estimate of $g_{m,k}[\ell]$ from $\bar{y}_{m,k}[\ell]$ is given by
\begin{equation}
    \hat{g}_{m,k}[\ell]=\frac{\sqrt{\rho\tau_c}\bar{\beta}_{m,k}}{\sqrt{\rho\tau_c}\bar{\beta}_{m,k}+\sigma^2}\bar{y}_{m,k}[\ell]
\end{equation}
where $\bar{\beta}_{m,k}=\sigma^2_{h_{m,k}}\beta_{m,k}$. 
Hence, the channel estimation error in CI $\ell$ is given by 
\begin{equation}
\tilde{g}_{m,k}[\ell]=g_{m,k}[\ell]-\hat{g}_{m,k}[\ell].
\end{equation}
Since $\tilde{g}_{m,k}[\ell]$ and $\hat{g}_{m,k}[\ell]$ are independent,
\begin{equation}
   \hat{g}_{m,k}[\ell]\backsim\mathcal{CN}(0,\gamma_{m,k}), 
\end{equation}
where $\gamma_{m,k}=\frac{\rho\tau_c\bar{\beta}_{m,k}^2}{\sqrt{\rho\tau_c}\bar{\beta}_{m,k}+\sigma^2}$
and $\tilde{g}_{m,k}[\ell]\backsim \mathcal{CN}(0,\bar{\beta}_{m,k}-\gamma_{m,k})$. Note that this estimation takes place locally at each AP.

\subsection{Downlink Transmission}
In TDD, the channels are reciprocal, and thus the estimated uplink channels can be used for downlink transmission beamforming.\footnote{Note that the uplink can be analyzed similarly.}  Each AP treats the estimated or the predicted channel as the true channel, and applies conjugate beamforming to transmit to the users. The transmitted signal from  AP $m$ in CI $\ell$ is given by 
\begin{equation}
    x_m[\ell]=\sqrt{p}_d\sum_{k=1}^K \sqrt{\eta_{m,k}} \hat{g}_{m,k}^*[\ell]s_k,
\end{equation}
where $p_d$ is the total downlink transmit power, $\eta_{m,k}$ is a power coefficient assigned at AP $m$ to user $k$, 
%and should be selected to achieve the total power constraint $\mathbb{E}\big[|x_m|^2\big]\leq p_d$, 
and $s_k$ is the codeword symbol intended for user $k$, which satisfies $\mathbb{E}\big[|s_k|^2\big]=1$. Normally, $\eta_{m,k}$ should be selected to optimize performance while satisfying the power constraint. However, since the power allocation problem is not within the scope of this paper, we assume that the power is allocated equally among the users, i.e., $\eta_{m,k}=\frac{1}{\sum_k \mathbb{E}\big[|\hat{g}_{m,k}|^2\big]}\  \forall k$. Therefore, the received signal at the user $k$ during CI $\ell$ is given by
\begin{equation}
    r_k[\ell]=\sum_{m=1}^M g_{m,k}[\ell]x_m[\ell]+n_k,
\end{equation}
where $n_k \backsim \mathcal{CN}(0,1)$ is a circularly symmetric complex Gaussian noise at the user $k$. In this paper, we consider the system's net throughput  as a performance metric for the proposed scheme and for baseline schemes. The performance metrics are presented in Section \ref{Sec:Perf}.
%\section{Baseline Approach: Conventional MMSE CE Approach}

%==========================================================Section=================================================
%At the $i$th resource block, the estimated channel of the user $k$ that belongs to the $n$th group is given by 
%\begin{equation}
%    \hat{\mathbf{g}}_{m,k}=\frac{\sqrt{\rho\tau_c}\beta_{m,k}}{\sqrt{\rho\tau_c}(\beta_{m,%k})+\sigma^2}\bar{y}_{m,k}
%\end{equation}
%where $\bar{y}_{m,k}=\sqrt{\rho\tau_c} g_{m,k}+\mathbf{q}_kn$. The channel of the users that do not belong to the group $n$ would be predicted at the $i$th block rather than estimated.
\section{Proposed CE and prediction (CEP) Scheme}

\begin{figure}
     \centering
     \begin{subfigure}[b]{0.45\textwidth}
         \centering
         \includegraphics[width=\textwidth]{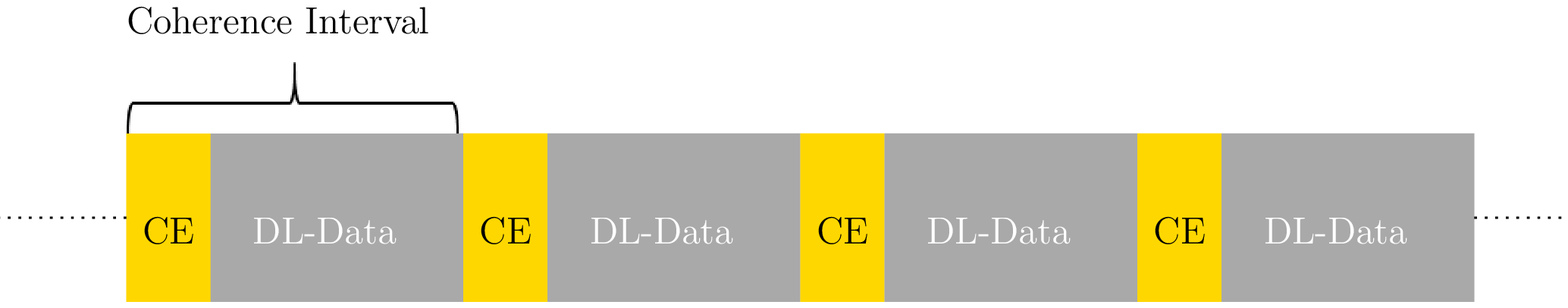}
         \caption{Traditional TDD scheme}
         \label{fig:TDD}
     \end{subfigure}
     \hfill
     \begin{subfigure}[b]{0.45\textwidth}
         \centering
         \includegraphics[width=\textwidth]{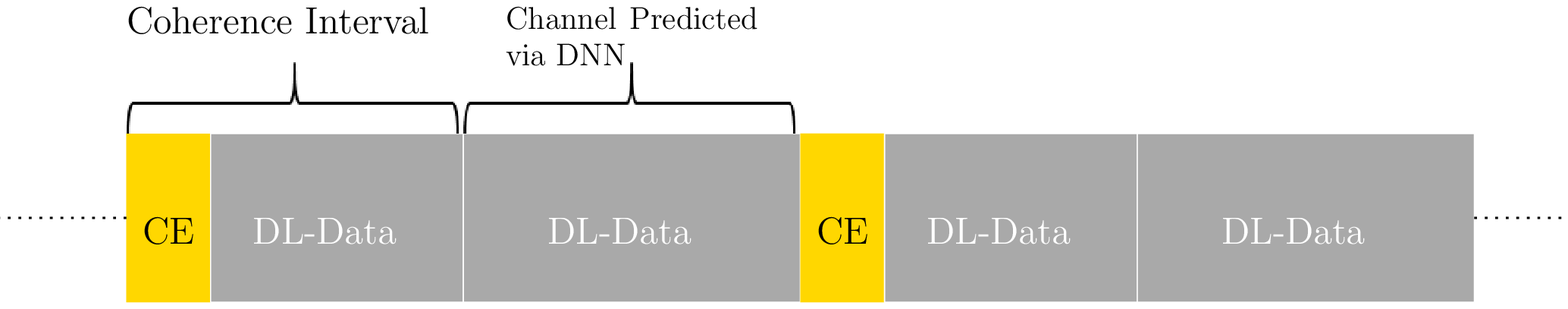}
         \caption{DL-based proposed scheme with 1:1 E2P ratio}
         \label{fig:DNN_model}
     \end{subfigure}
    \caption{Comparison between the proposed and the traditional TDD scheme}
    \label{fig:TDD_DNN}
\end{figure}
     
In this section, we describe the proposed hybrid CEP scheme and how the training input and output dataset is generated. When the number of users is large or the CI is short, sending pilot signals in all CIs leads to a large overhead compared to the transmitted data. On the other hand, predicting CSI by employing outdated estimated or a pre-predicted CSI leads to inaccurate CSI. Hence, this paper proposes to trade-off between estimation via pilot transmission and prediction. Here, we explain the proposed scheme under the assumption that the ratio of estimation to prediction (E2P) is 1:1, then we generalize for different ratios. 
\subsection{The CEP Scheme}
For a 1:1 E2P ratio, the proposed scheme alternatively estimates and predicts the CSI of consecutive CIs. In other words, at the beginning of CIs $\ell=2m+1,\ m=0,1,...$, which we call the estimate-and-transmit (ET) CIs, each user sends a pilot signal, then each AP uses the received signal to estimate the CSI of CI $\ell$ using an MMSE approach, which is used for beamforming in the downlink transmission.  Next, each AP employs the received pilot signals of the previous $\lceil\frac{Q}{2}\rceil$ ET CIs to predict the CSI of CI $\ell+1$, which we call the predict-and-transmit (PT) CI. The predicted CSI is then used for downlink beamforming in this PT CI.  This alternating procedure is repeated for all $\ell$.  This scheme and the classical TDD scheme are described in Fig. \ref{fig:TDD_DNN}. It can be seen that the CE overhead in the proposed scheme with a 1:1 E2P ratio is half that in the classical TDD scheme. 

This scheme can be extended to one ET CI followed by several PT CIs. For instance, for a 1:2 E2P ratio, each user sends the pilot signals every three CIs. APs then estimate the CSI of CI $3\ell$ (ET CI) and predict the CSI of CIs $3\ell+1$ and $3\ell+2$ (PT CIs). It is important to note that decreasing the E2P ratio reduces the CE overhead at the expense of a larger MSE. Hence, the E2P ratio must be selected carefully based on the given users' speed, number of users in the system, and the CI's duration.
\subsection{DNN-based Prediction}
In this paper, we propose to predict CSI using a DNN. We focus first on a 1:1 E2P ratio, then we generalize the result to different ratios. Without loss of generality, we assume that the CSI of CI $\ell=2m+1$ (ET CI) is estimated and the CSI of CI $\ell+1$ (PT CI) is  predicted. From equation \eqref{hmk}, we notice that the fading channel during CI $\ell$ is a weighted sum of the previous $Q$ channels. Therefore, we can use the received signals of the ET CIs that occur during the previous $Q$ CIs  to predict the CSI of the PI CI. In the proposed scheme, half of these previous CIs are estimated via MMSE and the remaining half are predicted. Since using the predicted half in addition to the estimated half leads to error propagation (which we observed in our numerical evaluations), we only use the estimated half as an input for the DNN. Therefore,  In our DNN, our goal is to predict the channel of CI $\ell+1=2m+2$ by exploiting the CSI of the previous CIs $\ell, \ell-2, \ell-4, \ldots, \ell-Q+2$ when $Q$ is even, or $\ell, \ell-2, \ell-4, \ldots, \ell-Q+1$ when $Q$ is odd, which are the ET CIs where channels are  estimated using MMSE. 
%This approach proposes that the channels of the odd CIs are estimated using the MMSE, while at the even CIs, we remove the CE overhead and predict the channels using deep learning approach. Fig. \ref{fig:TDD_DNN} compares between the proposed scheme and the traditional TDD scheme. In the TDD scheme, each AP estimates the users' channels at each CI by processing the received pilot signals. Whereas, in the proposed approach, each AP estimates the channel at every couple CI. The channel of the second CI is predicted using the previous gathered CSI.  

To create the dataset, we generate channels (ground truth) for each AP-user pair and for a large number of CIs. In order to allow  users to be mobile at different velocities, the dataset must include channels with various normalized Doppler shifts $f_n$. Here, there are two methods that can be followed when training the DNN. We can train a general DNN which learns to predict channels with different correlation properties. Alternatively, we can train several DNNs each of which is trained for a pre-defined range of $f_n$. We opted for the second method since the first showed lower 
 accuracy. Specifically, we trained three models, one for each of the intervals $f_n \in (0.05 \ \ 0.1]$,  $f_n \in [0.11 \ \  0.16)$, and  $f_n \in [0.16 \ \  0.2]$.  As we mentioned earlier, we assume a fixed CI's duration for all users, which means that each user has a different $f_n$. Note that as $f_n$ increases, the channel variation within CIs increases. To keep the assumption that the channel variation within each CI is negligible, we consider only the users whose Doppler shifts satisfy $f_n\leq f_{n,max}$, where $f_{n,max}$ is the maximum value of $f_n$ below which we can claim that the channel is nearly constant within a CI. For users whose $f_n > f_{n,max}$, it would be more appropriate to predict the channels' variations within CIs to reduce the channel aging effect. In Section \ref{Sec:Ext}, we discuss how the proposed scheme can be extended to predict the channel variations within CIs when $f_n > f_{n,max}$. Here, we select $f_{n,max}=0.2$ following \cite[Fig. 4]{CSI_Cor_Learning_2}  which shows that when $f_n>0.2$, channel aging during every CI reduces the downlink average achievable sum-rates significantly.   We also focus on $f_n > 0.05$ because if $f_n = 0.05$, it would be more appropriate to send a pilot once every two CIs for the MMSE and once every four CIs for the proposed 1:1 DL-based scheme, which leads to having $f_n = 0.1$. 
% in Fig. 4 show that when $f_n>0.2$, channel aging during every CI reduces the downlink average achievable sum-rates significantly.   We also consider that $f_n > 0.05$ because otherwise, it would be more appropriate to send a pilot once every two CIs for the MMSE and once every four CIs for the proposed 1:1 DL-based scheme, which leads to the same results when we have $f_n=0.1$.   
 
 %Note also that since the duration of the CI depends on $f_n$, to achieve the assumption that the channel within each CI is approximately fixed, we need to reduce the CI duration as the value of $f_n$ increases or passes a certain threshold. In other words, reducing the CI would keep the value of $f_n$ to be less than a certain threshold.    
% Authors of \cite{CSI_Cor_Learning_2}  in Fig. 4 show that when $f_n=0.2$, channel aging during every CI reduces the downlink average achievable sum-rates of by half. In this paper, we assume that $f_n \leq 0.2$. Note that reducing the CI's length would increase the percentage of overhead and this makes the proposed approach more powerful than the classical TDD scheme. However, we only consider the case that the CI length is fixed and $f_n \leq 0.2$.    

%the CI length must be increased to mitigate the effect of the CE overhead, where the channel aging during the CI is negligible (based on our numerical results).

\begin{figure}[!t]
\centering
\includegraphics[scale=0.5]{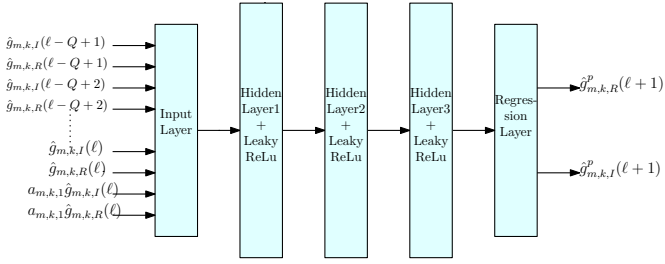}
\caption{DNN model with 1:1 E2P ratio, assuming Q is odd.}
\label{fig:DNN}
\end{figure}
As shown in Fig. \ref{fig:DNN}, we propose a multi-layer perceptron (MLP) neural network to predict the channel $g_{m,k}(\ell+1)$. To predict $g_{m,k}(\ell+1)$, the input of the DNN  consists of the real parts and the imaginary parts of the previously estimated channels $\hat{g}_{m,k,R}(\ell),\hat{g}_{m,k,I}(\ell), \hat{g}_{m,k,I}(\ell-2), \hat{g}_{m,k,I}(\ell-2),\ldots, \hat{g}_{m,k,R}(\ell-Q+m), \hat{g}_{m,k,I}(\ell-Q+m),$ where $m=2$ if $Q$ is even and $m=1$ if $Q$ is odd, and where the subscript ${\rm R}$ and ${\rm I}$ denotes the real an the imaginary parts, respectively. Additionally, we input  $ -a_{m,k,1}\hat{g}_{m,k,R}(\ell), -a_{m,k,1}\hat{g}_{m,k,I}(\ell)$, which tells the DNN how correlated the current and the previous CIs are, where $-a_{m,k,1}$ is a function of the normalized Doppler shift $f_n$ as shown in Section II. In addition, the output is also separated into the real and the imaginary parts.

As shown in Fig. \ref{fig:DNN}, the proposed DNN consists of an input layer, three hidden layers, and an output regression layer. The number of perceptrons at each hidden layer is $2^{10}$. The input and output size of the proposed DNN are $Q+2$ and $2$, respectively. Leaky ReLU activation functions are employed after each hidden layer. We use the adaptive moment estimation (ADAM) optimization technique to train the model parameters, using a mean square error (MSE) loss function. The learning rate is chosen to be $10^{-4}$, and the inputs and the outputs of the DNN are normalized to have zero mean and unity variance.   These DNN  specifications were selected since they lead to the smallest MSE among several options that were investigated.  

In a practical CF-mMIMO network, the DNN will be trained  offline at the CPU. The CPU then distributes these models to the APs to implement an online prediction relying on the received pilot signals from the users. Based on the users' mobility, each AP selects the appropriate model for each user. Note that similar to prediction, the estimation is done locally at each AP.  

The proposed model can be readily extended to a 1:N E2P ratio,  to predict $g_{m,k}(\ell+1), g_{m,k}(\ell+2),\ldots, g_{m,k}(\ell+N)$, for $\ell=(N+1)m+1$ and $m=1,2,....$ The inputs of the DNN  are the real and the imaginary parts of the previously estimated channels $\hat{g}_{m,k,R}(\ell),\hat{g}_{m,k,I}(\ell), \hat{g}_{m,k,R}(\ell-N), \hat{g}_{m,k,I}(\ell-N),\ldots, \hat{g}_{m,k,R}(\ell-Q), \hat{g}_{m,k,I}(\ell-Q+m),$ where $m=2$ if $Q$ is even and $m=1$ if $Q$ is odd. In addition, we input $ -a_{m,k,1}\hat{g}_{m,k,R}(\ell), -a_{m,k,1}\hat{g}_{m,k,I}(\ell)$. The output of the DNN now must be $2N$ dimensional, since the DNN must predict $2N$ values which are the real and the imaginary parts of the $N$ channels to be predicted.

 %To predict the channels $\ell+1, \ell+2,\ldots, \ell+N$, the input to the DNN must be $\ell, \ell-N-1, \ell-2\times N-1, \ldots, \ell-Q $. In addition, the output of the DNN must be $2N$, i.e., the DNN must be designed to predict $2N$ values which are the real and the imaginary values of the $N$ channels. 
%The loss function can be expressed as follows
%\begin{equation}
%    C_{DL}=\frac{1}{2}\sum
%\end{equation}

%\subsection{Creating the data set}

%Different from the proposed DL-based approaches in the literature, we here set two objectives which are estimate the channel of one CI and predict the channel of each another CI. In other words, we design DNN model where the output is two values; the first one is the estimated channel of the CI $i$ and the second one is the predicted channel of the CI $i+1$. 

%Increasing the number of orthogonal pilots leads to reducing the channel estimation error by minimizing the pilot contamination effect. However, this advantage achieved at the expense of the overhead. In other words, the problem of assigning time for channel estimation is a trade-off problem. On the other hand, machine learning based approaches have been proposed to predict the channels without consuming the time resource for channel estimation. Both above approaches have drawbacks and advantages, however combining both of them would lead to better performance.

%==========================================================Section=================================================
\subsection{Baseline Approach: Identity Mapping}
For comparison, we use a scheme similar to the proposed one but with prediction replaced by an identity mapping. In particular, the scheme shown in Fig. \ref{fig:DNN_model} is considered, where the predicted channel of an even-index CI equals to the estimated channel in the previous CI. The goal of comparing this approach with the proposed scheme is to prove that the DL-based prediction does not provide a trivial solution like an identity mapping.  
%This approach performs better when the normalized Doppler shift is small since the channel correlation between any consecutive CIs is very high. 

We also compare the proposed scheme with the TDD based approach (Fig. \ref{fig:TDD}), where channels are estimated every CI.  
%performs better when the number of users is low and the normalized Doppler shift is high.  
We demonstrate next that the proposed DL-based approach performs better than both approaches over all $f_n$.

%The goal of considering this approach as a baseline is to demonstrate that the proposed DNN approach does not provide a trivial solution such as the predicted channel equals to the previous estimated channel. 
%In other words, if the DNN-based approach provides the same performance as this baseline approach, 

\section{Performance Analysis and Simulation Results}
\subsection{Performance Analysis}
 \label{Sec:Perf}
%==========================================================Section=================================================

In this section, we present the performance metrics that evaluate the proposed and the baseline approaches. 
%Here, we consider only the downlink spectral efficiency, where the extension to the uplink transmission is straightforward. 
The users are assumed to only have access to the channel statistics. In this case, the following achievable rate can be used to evaluate the proposed system \cite{Cell_Ngo}
\begin{equation}
    R_k[\ell]= \log_2 \bigg(1+\frac{|D_k|^2}{E\{|B_k|^2\}+\sum_{k'\neq k}^K E\{|U_{kk'}|^2\}+1}\bigg),
\end{equation}
where 
$$D_k=\sqrt{p_d}E\bigg[\sum_{m=1}^M\sqrt{\eta_{m,k}}g_{m,k}[\ell]\bar{g}_{m,k}[\ell]\bigg],$$
\begin{multline*}
B_k=\sqrt{p_d}\bigg(\sum_{m=1}^M\sqrt{\eta_{m,k}}g_{m,k}[\ell]\bar{g}_{m,k}[\ell]\\
-E\bigg[\sum_{m=1}^M\sqrt{\eta_{m,k}}g_{m,k}[\ell]\bar{g}_{m,k}[\ell]\bigg]\bigg),
\end{multline*}
 and
$$U_{kk'}=\sqrt{p_d}\sum_{m=1}^M\sqrt{\eta_{m,k}}g_{m,k}[\ell]\bar{g}_{m,k}[\ell],$$
where $\bar{g}_{m,k}[\ell]$ is the estimated channel $\hat{g}_{m,k}$ in an ET CI, the predicted one in a PT CI, or the same channel as in the previous CI when an identity mapping is used for prediction. 
%\begin{equation}
%    \bar{g}_{m,k}[\ell]=\begin{cases} \hat{g}_{m,k}[\ell], & \text{ if CI } \ell \text{ is estimated};\\
%    g_{m,k}^p[\ell] & \text{ if CI } \ell \text{ is predicted},
%    \end{cases}
%\end{equation}
 Denote by $R_k^{CEP}[\ell]$ the achievable rate of the proposed CEP scheme, $R_k^{IM}[\ell]$ the achievable rate of the identity mapping approach, and by $R_k^{TDD}[\ell]$  the achievable rate of the classical TDD scheme. By taking into account the channel estimation overhead, the net throughput of the system using the CEP approach is given by 
\begin{equation}
    S^{CEP}=B(1-\frac{\alpha \tau_c}{(1+\alpha)\tau}) \sum_{k=1}^K R_k^{CEP},
\end{equation}
where $\alpha$ is the E2P ratio (e.g. if the E2P ratio is 1:N, $\alpha =\frac{1}{N}$).  The net throughput of the identity mapping approach with $\alpha$ E2P ratio is given by
\begin{equation}
    S^{IM}=B(1-\frac{\alpha \tau_c}{(1+\alpha)\tau}) \sum_{k=1}^K R_k^{IM}.
\end{equation}
 The net throughput of the system using the TDD scheme is given by
\begin{equation}
    S^{TDD}=B(1- \tau_c/\tau) \sum_{k=1}^K R_k^{TDD}.
\end{equation}

 \subsection{Simulation Results}
 \label{SResult}

%==========================================================Section=================================================
\begin{figure}[!t]
\centering
\includegraphics[width=3in]{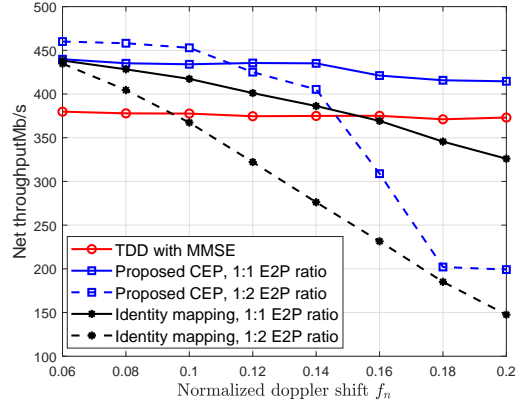}
\caption{Net throughput versus normalized Doppler shift $f_n$ for $M = 64$ and $K = 40$.}
\label{R_fn}
\end{figure}

In this section, we aim to evaluate the proposed CEP approach and compare it to the baseline approaches. We deploy a $64$ distributed APs (except in Fig. \ref{R_M}) over an area of $1$ Km$^2$. For every user-AP link, we generate $150$ consecutive CIs under the assumption that $Q=63$. The large scale fading $\beta_{m,k}$ expression given in \cite{Cell_Ngo} is adopted in our simulation. The carrier frequency and the bandwidth are chosen to be $1.9$ GHz and $20$ MHz, respectively. The power assigned for pilot signals and data transmission are given respectively as $\rho=p_d=0.1$ Watt. The noise power spectral density is assumed to be $-122$ dB/Hz. 
%All the results points are an average of the results with 100 different APs and users distributions. 
In the CEP approach, we generate training data with different values of $f_n$ (uniformly distributed between 0.05 and 0.2) and train models offline. Then test data is generated based on the transmission of pilot signals for various  users and AP locations.  
\begin{figure}[!t]
\centering
\includegraphics[width=3in]{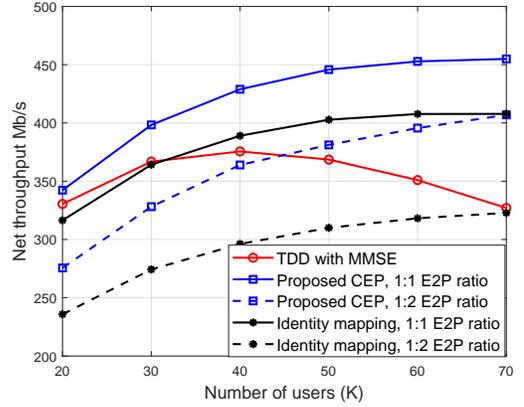}
\caption{Net throughput versus number of users for $M = 64$.}
\label{R_K}
\end{figure}

Fig. \ref{R_fn} shows the effect of the normalized Doppler shift $f_n$ on the net throughput. In this figure, we consider the 1:1 and 1:2 E2P ratios. First, the figure shows that the channel prediction accuracy decreases with increasing $f_n$ since increasing $f_n$ decreases the channel correlation between consecutive CIs. The proposed CEP scheme with 1:1 E2P ratio outperforms the TDD approach, the 1:1 and 1:2 identity mapping approach, for all considered values of $f_n$. It can also be seen that for low-mobility users where $f_n$ is less than $\approx 0.12$, an E2P ratio of 1:2 performs better than 1:1 due to the lower overhead and CSI correlations remain high for more than two CIs.  In all cases, the proposed CEP demonstrates its superiority over the baseline approaches. 
%\begin{equation}
%    \beta_{m,k}=
%\end{equation}

\begin{figure}[!t]
\centering
\includegraphics[width=3in]{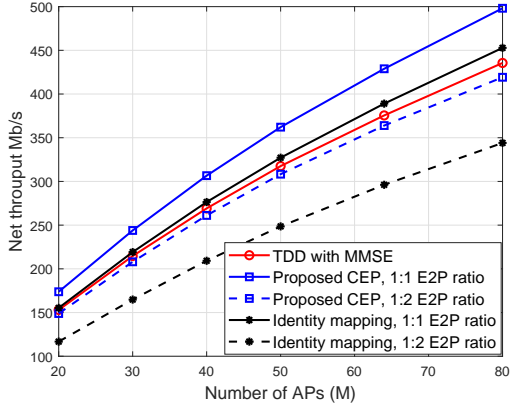}
\caption{Net throughput versus number of APs for $K = 40$.}
\label{R_M}
\end{figure}

A major factor that exacerbates the CE overhead is the large number of users. Fig. \ref{R_K} shows how increasing the number of users affects the net throughput. In this figure, we assume that the users have a random velocity, so $f_n$ is uniformly distributed in the interval $[0.05\ 0.2]$. The figure shows that for the TDD-MMSE scheme, the rate increases until some point, then it decreases when the CE overhead starts to dominate. It can be seen that the DL-based proposed approach with 1:1 ratio outperforms the TDD-MMSE and the identity mapping approaches.

In cell-free systems, the number of APs is expected to be massive. Hence, Fig. \ref{R_M} shows how increasing the number of AP would affect the net throughput. It can be seen that the DL-based approach works better than the TDD-MMSE whether the number of APs is small or large. 

\section{Extension the Proposed Scheme to Solve Other Problems}
\label{Sec:Ext}
\textbf{Channel Aging Within CIs:}  The authors of \cite{CSI_Cor_Learning_2} and \cite{zheng2021impact} showed that channel aging within CIs has a significant impact on the channel estimation error. The proposed hybrid CEP scheme can be extended to mitigate the impact of the channel aging within CIs  to solve this problem. In particular, each CI can be divided into $N+1$ sub-intervals, where the channel is estimated in the first sub-interval and predicted in the remaining $N$ sub-intervals using the proposed DNN.  This process can be repeated in every CI. 

\textbf{Fixed CE Time:} Some standards (e.g., in LTE) suggest that the assigned time of the uplink training phase must be fixed for each CI, where the training overhead is for example $1\%$ of the CI time. The proposed scheme can be extended to consider such a case. Particularly, if the number of users is $K$ and the number of orthogonal pilots is $M < K$, we can schedule all user so that in each CI, only $M$ users (or less) send pilot signals to have their channels estimated, while the channels of the remaining users can be predicted using previously estimated channels. Then, the set of users whose channels are estimated changes from one CI to the next so that each users' channel is estimated at least one every $\lceil \frac{K}{M}\rceil$ CIs.   For example, suppose that we have $9$ users and the number of orthogonal pilots is $5$. During the CI $\ell$,  the first five users send pilot signals so that their channels can be estimated, while the last four users' channels will be predicted. During CI $\ell+1$, we flip the process, where the last four users send pilot signals so that their channels can be estimated, while  the first five users' channels will be predicted. This process then can be repeated every two CIs for a 1:1 E2P ratio for all users.

\section{Conclusion}
This paper considered the problem of CE overhead in cell-free massive MIMO systems and proposed a DL-based alternating channel estimation and prediction scheme to reduce its effect. The proposed scheme is based on exploiting the temporal correlation to predict the channel of every second CI.  Simulation results showed that the number of users and the user mobility play an important role in increasing the CE overhead and hence the system throughput. The results also showed that the proposed DL-based scheme reduced the CE overhead, and hence improved the system throughput over the classical TDD-based approach.

\bibliographystyle{IEEEtran}
\bibliography{main}

\begin{thebibliography}{10}
\providecommand{\url}[1]{#1}
\csname url@samestyle\endcsname
\providecommand{\newblock}{\relax}
\providecommand{\bibinfo}[2]{#2}
\providecommand{\BIBentrySTDinterwordspacing}{\spaceskip=0pt\relax}
\providecommand{\BIBentryALTinterwordstretchfactor}{4}
\providecommand{\BIBentryALTinterwordspacing}{\spaceskip=\fontdimen2\font plus
\BIBentryALTinterwordstretchfactor\fontdimen3\font minus
  \fontdimen4\font\relax}
\providecommand{\BIBforeignlanguage}[2]{{%
\expandafter\ifx\csname l@#1\endcsname\relax
\typeout{** WARNING: IEEEtran.bst: No hyphenation pattern has been}%
\typeout{** loaded for the language `#1'. Using the pattern for}%
\typeout{** the default language instead.}%
\else
\language=\csname l@#1\endcsname
\fi
#2}}
\providecommand{\BIBdecl}{\relax}
\BIBdecl

\bibitem{Cell_Ngo}
H.~Q. Ngo, A.~Ashikhmin, H.~Yang, E.~G. Larsson, and T.~L. Marzetta,
  ``Cell-free massive {MIMO} versus small cells,'' \emph{IEEE Trans. Wireless
  Commun.}, vol.~16, no.~3, pp. 1834--1850, 2017.

\bibitem{dong2019deep}
P.~Dong, H.~Zhang, G.~Y. Li, I.~S. Gaspar, and N.~NaderiAlizadeh, ``Deep
  {CNN}-based channel estimation for {mmWave} massive {MIMO} systems,''
  \emph{IEEE J. Sel. Topics Signal Process.}, vol.~13, no.~5, pp. 989--1000,
  2019.

\bibitem{kim2018deep}
K.~Kim, J.~Lee, and J.~Choi, ``Deep learning based pilot allocation scheme
  ({DL-PAS}) for {5G} massive {MIMO} system,'' \emph{IEEE Commun. Lett.},
  vol.~22, no.~4, pp. 828--831, 2018.

\bibitem{jiang2020deep}
W.~Jiang and H.~D. Schotten, ``Deep learning for fading channel prediction,''
  \emph{IEEE Open J. Commun. Society}, vol.~1, pp. 320--332, 2020.

\bibitem{8482358}
T.~Wang, C.-K. Wen, S.~Jin, and G.~Y. Li, ``Deep learning-based {CSI} feedback
  approach for time-varying massive {MIMO} channels,'' \emph{IEEE Wireless
  Commun. Lett.}, vol.~8, no.~2, pp. 416--419, 2019.

\bibitem{Mash_Pruning}
M.~B. Mashhadi and D.~Gündüz, ``Pruning the pilots: Deep learning-based pilot
  design and channel estimation for {MIMO-OFDM} systems,'' \emph{IEEE Trans.
  Wireless Commun.}, pp. 1--1, 2021.

\bibitem{zhang2020deep}
Y.~Zhang, M.~Alrabeiah, and A.~Alkhateeb, ``Deep learning for massive {MIMO}
  with 1-bit {ADCs}: When more antennas need fewer pilots,'' \emph{IEEE
  Wireless Commun. Lett.}, vol.~9, no.~8, pp. 1273--1277, 2020.

\bibitem{yuan2020machine}
J.~Yuan, H.~Q. Ngo, and M.~Matthaiou, ``Machine learning-based channel
  prediction in massive {MIMO} with channel aging,'' \emph{IEEE Trans. Wireless
  Commun.}, vol.~19, no.~5, pp. 2960--2973, 2020.

\bibitem{ARC}
K.~E. Baddour and N.~C. Beaulieu, ``Autoregressive modeling for fading channel
  simulation,'' \emph{IEEE Trans. Wireless Commun.}, vol.~4, no.~4, pp.
  1650--1662, 2005.

\bibitem{CSI_Cor_Learning_2}
K.~T. Truong and R.~W. Heath, ``Effects of channel aging in massive {MIMO}
  systems,'' \emph{J. Commun. and Net.}, vol.~15, no.~4, pp. 338--351, 2013.

\bibitem{box2015time}
G.~E. Box, G.~M. Jenkins, G.~C. Reinsel, and G.~M. Ljung, \emph{Time series
  analysis: forecasting and control}.\hskip 1em plus 0.5em minus 0.4em\relax
  John Wiley \& Sons, 2015.

\bibitem{zheng2021impact}
J.~Zheng, J.~Zhang, E.~Bj{\"o}rnson, and B.~Ai, ``Impact of channel aging on
  cell-free massive mimo over spatially correlated channels,'' \emph{IEEE
  Trans. Wireless Commun.}, vol.~20, no.~10, pp. 6451--6466, 2021.

\end{thebibliography}

\end{document}